\documentclass[12pt]{article}

\usepackage{sbc-template}
\usepackage{graphicx,url}
\usepackage[utf8]{inputenc}

\newcommand{\code}[1]{\texttt{#1}}
\title{Towards Jacamo-rest: A Resource-Oriented Abstraction for Managing Multi-Agent Systems\thanks{This study was partially funded by Project AG-BR of Petrobras and by the program PrInt CAPES-UFSC ``Automação 4.0'' and partially funded by the Wallenberg AI, Autonomous Systems and Software Program (WASP).}}

\author{Cleber J. Amaral\inst{1,2}, Jomi F. Hübner\inst{1} and Timotheus Kampik\inst{3}}

\address{Universidade Federal de Santa Catarina (UFSC)\\
  Florian{\'o}polis -- SC -- Brazil
\nextinstitute
  Instituto Federal de Santa Catarina (IFSC)\\
  S{\~a}o Jos{\'e} -- SC -- Brazil
\nextinstitute
  Umeå University\\
  Umeå -- Sweden
  \email{\vbox{\noindent cleber.amaral@ifsc.edu.br, jomi.hubner@ufsc.br and tkampik@cs.umu.se}}
}

\begin{document} 

\maketitle

\begin{abstract}
The Multi-Agent Oriented Programming (MAOP) paradigm provides abstractions to model and implements entities of agents, as well as of their organisations and environments.
In recent years, researchers have started to explore the integration of MAOP and the resource-oriented web architecture (REST).
This paper further advances this line of research by presenting an ongoing work on \textit{jacamo-rest}, a resource-oriented web-based abstraction for the multi-agent programming platform \textit{JaCaMo}.
\textit{Jacamo-rest} takes Multi-Agent System (MAS) interoperability to a new level, enabling MAS to not only interact with services or applications of the World Wide Web but also to be managed and updated in their specifications by other applications.
To add a developer interface to \textit{JaCaMo} that is suitable for the Web, we provide a novel conceptual perspective on the management of MAOP \textit{specification entities} as web resources.
We tested \textit{jacamo-rest} using it as a middleware of a programming interface application that provides modern software engineering facilities such as continuous deployments and iterative software development for MAS.

\end{abstract}

\section{Introduction}

Multi-Agent Oriented Programming (MAOP) is an approach that uses first-class abstractions to design and develop distinct dimensions that characterise an MAS, among them agents, environments, and organisations~\cite{BOISSIER2013747}. The \textit{agent} dimension provides abstractions for the development of autonomous entities responsible for decision-making processes, called agents. The \textit{environment} dimension uses abstractions for the definition and programming of environmental artefacts, i.e., distributed non-autonomous resources that can be used and shared by agents to achieve their goals. Finally, the \textit{organisation} dimension allows structuring, coordinating, and regulating the interactions in the system, among agents, as well as between agents and the shared environment.

The Web, in contrast, has its abstractions specifically designed to sustainable support the Internet at scale. On the Web, interactions among applications and services frequently use REpresentational State Transfer (REST) Application Programming Interfaces (APIs)~\cite{6827107}. REST refers to an architecture that models the elements of the Web as resources~\cite{Fielding2002}.

Recent works have already explored the intersection of the Web, REST, and MAOP~\cite{Ciortea2018}, focusing on the integration of MAS and Web abstractions to facilitate run-time interaction between MAS and traditional web applications.
In particular, web service abstractions that allow applications to interact with Jason agents~\cite{10.1007/978-3-030-45778-5_28}, as well as with agents written in ASTRA~\cite{10.1145/3308560.3316509}, have been developed.
While the MAS integration with other applications is broadly covered in the literature, the support for the development and run-time modification of MAS specifications and code has so far not been considered. In this paper, we call this type of integration \textit{management integration}.
In particular, the paper presents an early version of the \textit{jacamo-rest} implementation that places MAS on the Web as a resource-oriented interface that supports management integration.

This paper is organised as follows: Section~\ref{mas} presents MAS entity abstractions, whereas Section~\ref{rest} presents the Web abstractions counterpart. Then, Section~\ref{jacamorest} introduces \textit{jacamo-rest}, an interface implementation for MAS on the Web.
In particular, the section highlights the concepts of \textit{management} and \textit{interaction} integration and presents the results of an early evaluation. Subsequently, Section~\ref{discussion} discusses the presented work, before Section~\ref{future-work} presents future research directions and Section~\ref{conclusion} concludes the paper.

\section{The Multi-Agent Oriented Model}\label{mas}

\begin{figure}[ht]
	\centering
	\includegraphics[width=0.8\textwidth]{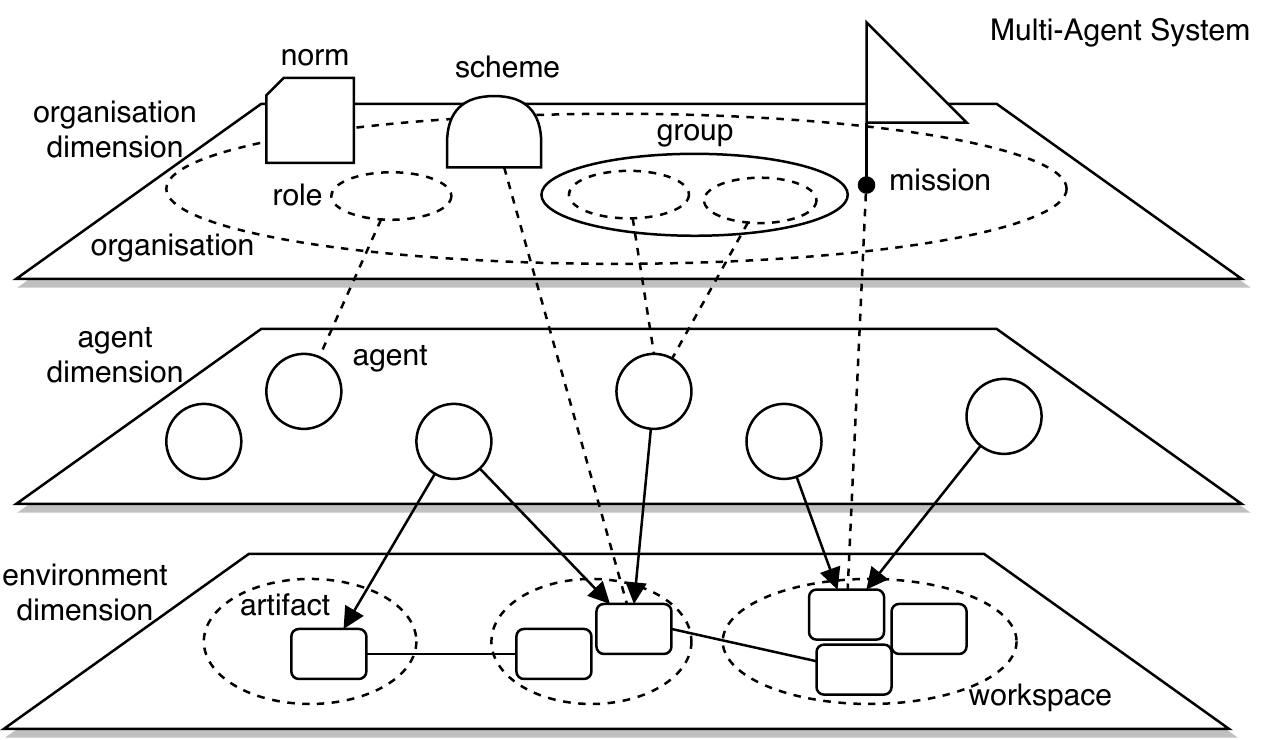}
	\caption{The JaCaMo multi-agent oriented programming model, adapted from~\cite{BOISSIER2013747}.}
	\label{fig:jacamo-model}
\end{figure}

The multi-agent oriented programming approach proposes a distinction of entities in dimensions of programming. The division in dimensions facilitates the development of entities of each class, making use of specific abstractions, languages and paradigms. In their work, \cite{BOISSIER2013747,boissier2019dimensions} propose three dimensions: (i) the environment dimension, (ii) the organisation dimension, and (iii) the agent dimension. Figure~\ref{fig:jacamo-model} illustrates these dimensions and relationships between agents and environmental and organisational artefacts.

The dimension of the environment is composed of \textit{artefacts} and \textit{workspaces}. Artefacts are virtual elements with observable properties and operations that can be triggered by agents~\cite{Ricci2011}. They represent non-autonomous entities that agents can use and share. Some examples of entities that can be modelled as artefacts are physical devices, virtual and physical resources such as non-autonomous software, tools and all sort of objects. An example of a physical device is a machine that has \textit{displays} for showing its state (observable properties) and buttons (operations). When an artefact represents an external entity, it usually has a mean to communicate with this concrete entity and mirrors this object, which allows agents to observe and control it just as if they could control the corresponding real-world object. \textit{Workspaces} are virtual boundaries for artefacts, working similarly to physical boundaries for real-world objects. An agent has to enter a workspace to observe and control artefacts within its boundaries.

The \textit{organisation} dimension is composed of organisational artefacts that describe abstract entities of organisations. Organisations can be an MAS's internal or external entities. Some organisational artefacts describe how agents are structured in groups, others define sub-goals schemes that must be achieved to fulfil a global goal, and yet others represent missions associated with sub-goals and organisational roles~\cite{Hubner2010}. These artefacts help maintain coherence among agents and usually provide means to coordinate them.

Finally, the \textit{agent} dimension is composed of autonomous entities that are situated in an environment, perceiving and acting in it. An artificial agent is a computer program with distinguishing features, of which we highlight: (i) the agent is autonomous, \textit{i.e.}, takes its own decisions including rejecting orders from other agents, even from humans; (ii) the agent is proactive, \textit{i.e.}, it is goal-oriented, which in practice means it has initiative and for chasing a goal it may perform several different plans; (iii) an agent is social, which means it is aware of the existence of other agents and can act in a cooperative way towards mutual goals~\cite{Wooldridge2002}.
Among possible actions that agents may perform we have: entering workspaces, reading observable properties of artefacts, operating artefacts, adopting organisational roles, fulfilling individual and collective goals, and communicating with other agents.

\section{Resource-Oriented Architecture and Representational State Transfer}\label{rest}

With the emergence of the Web as the world's most important application ecosystem, a new generation of software architecture paradigms have risen to prominence.
One of those paradigms is \textit{resource-oriented architecture} and in particular REpresentational State Transfer (REST)~\cite{fielding2000}, which is--at least so some extent--reflected in most web applications.
In REST, any information object that can be named, for example, an image, a document, or the virtual presentation of a physical object, is considered a resource~\cite{fielding2000}.
REST in a technology-agnostic protocol that specifies how such resources should be exchanged between a server and a client.

According to REST, each resource on a server has a Unique Resource Identifier (URI). Clients can create, retrieve, update, and delete resources on a server by sending requests against this server, whereby existing resources are accessed and manipulated via their URI.
When retrieving data from a resource type's endpoint, all resources of the corresponding type will be listed, or a subset of these resources, if a filter is applied by specifying query parameters.
REST aims to be stateless in that all state about a client's session is maintained by the client and not by the server.
Typically, REST is used in conjunction with the Hypertext Transfer Protocol (HTTP).
Then, REST can make use of HTTP verbs (\code{GET}, \code{POST}, \code{PUT}, \code{DELETE}, \code{OPTIONS}, \textit{et cetera}) to access and manipulate resources.
REST does not specify a data encoding for resources; the format can be, for example, HTML, XML, or JSON. \textit{RESTful services} are modelled using resource-oriented architecture (ROA), make proper use of HTTP verbs, and allow their Application Interfaces (APIs) to be discovered by other systems~\cite{richardson2008restful}. 

\section{The \textit{jacamo-rest} middleware}\label{jacamorest}

The JaCaMo platform\footnote{\url{http://jacamo.sourceforge.net/}.} joins different \textit{frameworks} for developing distinct dimensions of MAS. The agents that run on this platform inherit facilities from the Jason \textit{framework}, allowing the interpretation of the \textit{AgentSpeak} language, which includes among other functionalities sending messages among agents. Although this feature is a powerful tool, this integration is restricted to agents that run on the same platform instance. Among extensions of the JaCaMo ecosystem, we have the possibility to use JADE~\cite{Bellifemine2003} as an infrastructure. In this case, the JADE framework serves as an integration middleware for agents. However, this integration is also limited to artificial agents, which in this case can be Jason agents or JADE agents.

Another way to address integration in JaCaMo is through artefacts. An artefact can be used for integrating an agent and a non-autonomous external entity, and for integrating agents themselves. In the former case, an artefact can be built to communicate with an external entity embedding a specific protocol and virtualising the external entity, which allows agents to manipulate this representation. In the latter case, an artefact can be used as a means for communication among agents, for instance, a telephone or an email account artefact. Such artefacts may support communication between internal agents and external autonomous entities, for example, with humans. The limitation of this approach is that each kind of external entity may require the development of a specific integration.

\begin{figure}[ht]
	\centering
	\includegraphics[width=0.6\textwidth]{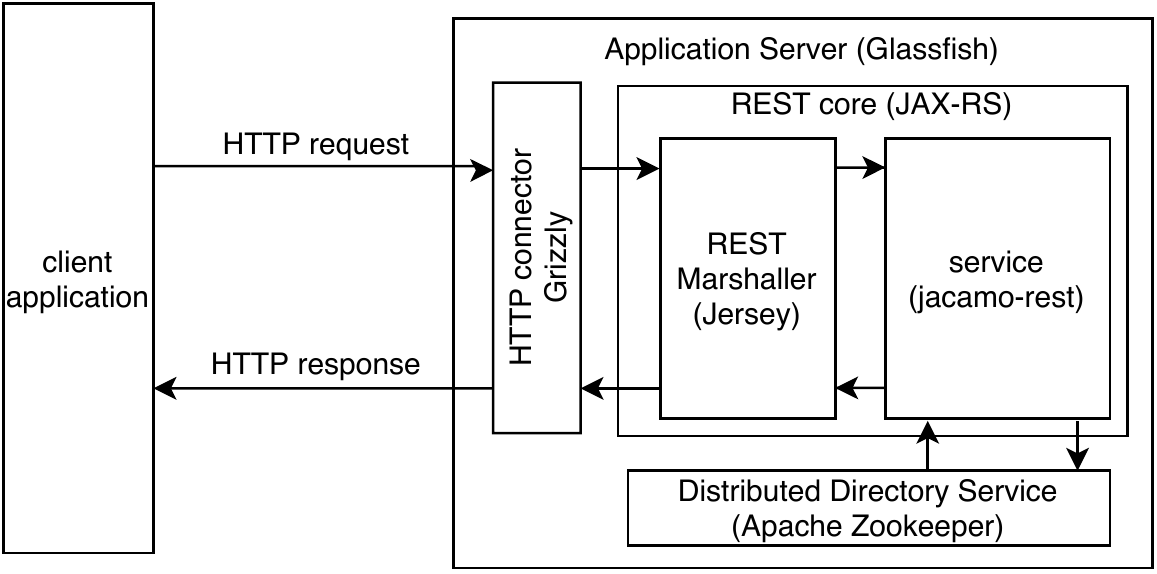}
	\caption{Data flow among libraries and frameworks used by \textit{jacamo-rest}}
	\label{fig:rest-overallarch}
\end{figure}

To expose JaCaMo entities as resources on the Web, we introduce \textit{jacamo-rest}. The \textit{jacamo-rest} extension\footnote{Jacamo-rest is an open-source project available at \url{https://github.com/jacamo-lang/jacamo-rest}.} is developed in Java, the same technology that the JaCaMo platform is based on. To support REST functionalities, \textit{jacamo-rest} adds libraries for defining endpoints and web infrastructure for providing web-server facilities. Figure~\ref{fig:rest-overallarch} illustrates the main libraries used in the extension and how data flows between them.

\textit{Jacamo-rest} is instantiated by the \textit{Glassfish} application server. The client, when communicating with \textit{jacamo-rest}, makes requests over the HTTP protocol. The \textit{Grizzly} HTTP connector receives and sends data in this protocol, executing low-level functionalities for mediating and routing HTTP messages. The connector delivers the content of the messages, \textit{i.e.}, the payload, to the REST \textit{Jersey} marshaller. \textit{Jersey} forwards the data, which results in method invocations defined in \textit{jacamo-rest} classes. Each method follows the specification of the \textit{JAX-RS} Java library, which defines how data enters and leaves the REST implementation. In other words, each REST method answers to an \textit{endpoint} provided by \textit{jacamo-rest}. Additionally, \textit{jacamo-rest} provides a directory service facility that publishes the list of agents and their functionalities similar to a distributed yellow-pages service.

\textit{Jacamo-rest} makes available mainly three collections of \textit{endpoints}: (i) related to agents (\code{/agents}), (ii) related to the environment (\code{/workspaces}), and (iii) related to organisations (\code{/organisations}). Each one of these collections is defined in a specific class that implements REST methods. For instance, the class \textit{RestImplAg} implements the \textit{endpoint} \code{GET /agents} that returns the list of agents and the \textit{endpoint} \code{GET /agents/\{agentuid\}} returns data about the agent identified by \code{agentuid}. 

\begin{figure}[ht]
\centering
\includegraphics[width=0.9\textwidth]{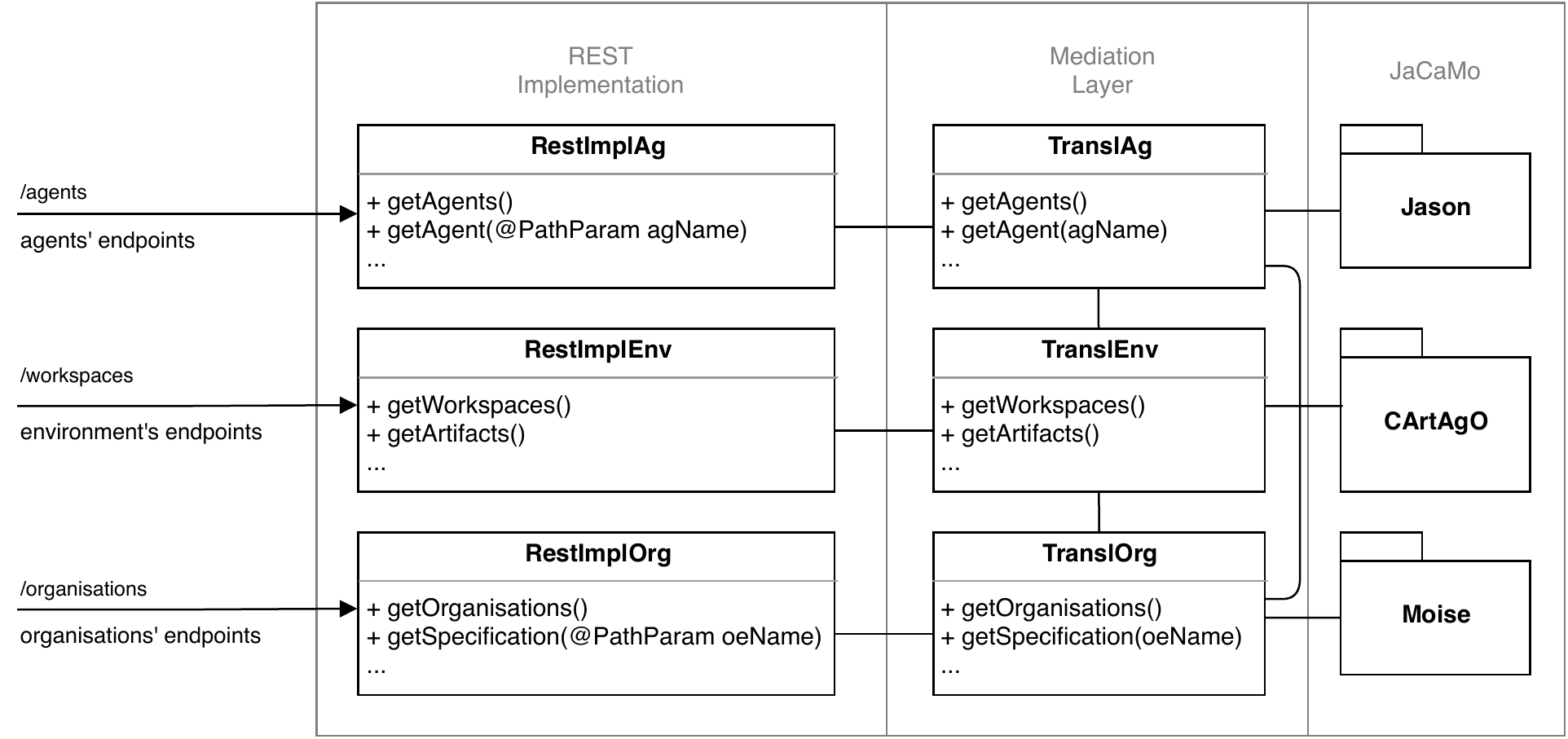}
\caption{Jacamo-rest's collections of endpoints and software layers view}
\label{fig:rest-internal}
\end{figure}

\textit{Jacamo-rest} has implementations for mediating data that flows between REST implementation classes and the JaCaMo platform. These mediation classes are responsible for delivering data from JaCaMo in the format that will be sent to the client through a REST interface. The mediation classes communicate among each other, which allows for the reuse of some methods. For example, the mediator of the environmental dimension has methods to format the data of artefacts. When an agent's endpoint needs to show artefact information, for instance, the artefacts an agent is observing, the mediator of agents uses data of the mediator of the environment to produce the final content. Figure~\ref{fig:rest-internal} illustrates how these classes are interconnected.

\subsection{Interaction Integration and Management Integration}
\label{integration}

Existing works primarily shed light on the \textit{interaction integration} of agents and multi-agent systems on the Web~\cite{Ciortea2018,10.5555/3306127.3331893,10.1007/978-3-030-45778-5_28,10.1145/3308560.3316509}; \textit{i.e.}, the works propose ways for agents and multi-agent systems to interact at run-time via the Web.
The purpose of RESTful APIs is, in this context, the provision and update of \textit{application state}.
Interaction integration is a requirement from the execution process view, \textit{i.e.}, it is concerned with enabling communications between an MAS and third-party systems according to the requirements of the systems' application domain.

In our approach, the integration among applications is provided by message exchange, which is the standard way agents interact with each other. Indeed, the interaction integration view merely requires access to the services the MAS can provide, which translates to write access to send messages to agents' mailboxes (\code{POST}).
It is important to mention that the \code{/inbox} endpoint is only suitable for inbound requests, \textit{i.e.}, the MAS working as a server. In Section~\ref{evaluation}, we point to our previous work that allows exposing MAS as clients as well.
It is also worth noting that messages to an agent's inbox can be considered a Remote Procedure Call (RPC) style functionality; it does not retrieve or manipulate particular resources; instead, it triggers agent actions.
A concern with this approach is its vulnerability on overloading the system with a possible flood of requisitions~\cite{richardson2008restful}.
In the context of REST, RPCs are considered anti-patterns, which triggers the question of how these JaCaMo API endpoints can be exposed in a more idiomatic manner.
We touch upon this issue in the Section~\ref{discussion}. Still about interactions, one may point out that \textit{jacamo-rest} is not exposing environmental and organisational artefacts. Indeed, currently, we are considering them as internal MAS representations. In this sense, if needed, agents should mediate the sharing of such resources.

\begin{figure}[!htb]
	\centering
	\includegraphics[width=1.0\textwidth]{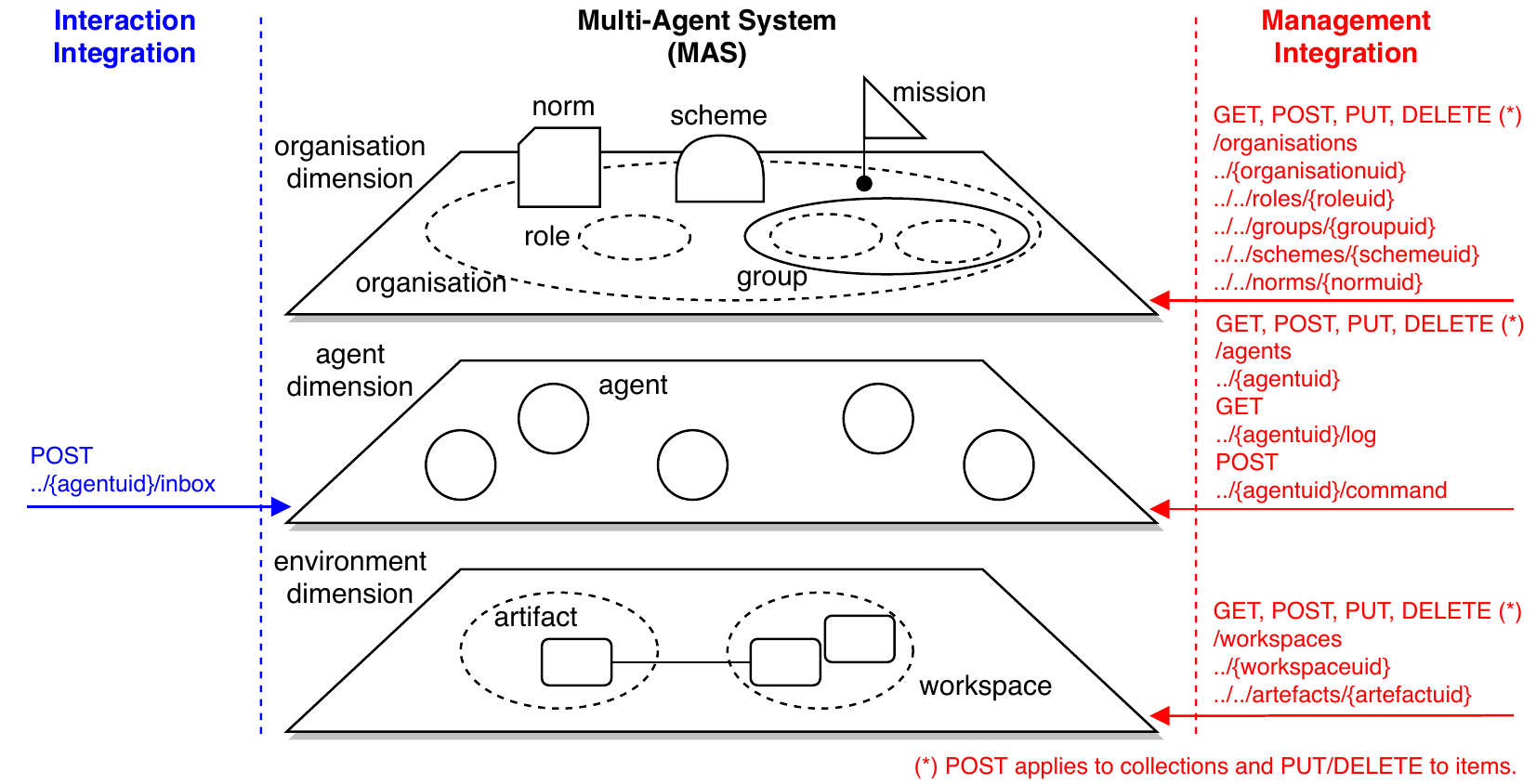}
	\caption{Endpoints and methods (HTTP verbs) shared between external applications and the MAS (left, blue), and for the MAS development (right, red).}
	\label{fig:jacamo-rest-integration}
\end{figure}

The notably facilities of \textit{jacamo-rest} is on \textit{management integration}, \textit{i.e.}, the update and retrieval of \textit{models} of agents, artefacts and their environment. Management integration is concerned with the development and operations view, and can be seen in the light of the \textit{DevOps} best practices in modern software development~\cite{7458761}, which attempt to facilitate the creation of autonomous development teams that are in charge of the whole process from development via testing and deployment to operations, and make use of a sophisticated and highly automated toolchain. Management integration requires a superset of the API endpoints that the interaction integration requires, because read and write access to both application state and models is important to deploy, update, debug and hot-fix the MAS and its models.\footnote{Our study does not take into account security issues.} Consequently, the approach is well-aligned with modern software engineering best practices. This new perspective further facilitates interoperability, distribution, and agility of MAS, as it allows developers to update MAS on the fly, and to automate these updates via continuous integration.
Figure~\ref{fig:jacamo-rest-integration} provides an overview of the resources and methods that are shared between application and management view (on the left-hand side), as well as the of ones that are specific to the management view (on the right-hand side).

Most \textit{management integration} endpoints are compliant with RESTful design best practices. MAS entities are properly modelled as resources, allowing operations such as retrieve, create, update, and delete resources. HTTP verbs are used as suggested, \textit{i.e.}, \code{GET} only retrieves data retrieve (and is safe and idempotent), \code{PUT} and \code{DELETE} are both idempotent, and \code{POST} creates, in most scenarios, new resources. The exception is the \code{POST} \code{/command} which is, just like the \code{/inbox} endpoint, an RPC-like method.

Finally, our implementation allows navigating application state via its API, in accordance with the \textit{Hypermedia As the Engine of Application State} (HATEOAS) concept. To do so, we use the Jersey facility that implements the \code{OPTIONS} verb for each exposed endpoint. This operation provides to the client all endpoints and possible methods that can be triggered. Developers and artificial agents can take advantage of this information and use it as a manual that helps them find resources and facilities provided by the API.

\subsection{Application Example}\label{evaluation}

We tested the \textit{jacamo-rest} concept of how MAS abstractions can be modelled as resources by implementing the \textit{jacamo-web} application\footnote{Jacamo-web is available at \url{https://github.com/jacamo-lang/jacamo-web}.}~\cite{amaral2019jacamo}. The \textit{Jacamo-web} Integrated Development Environment (IDE) allows programming MAS interactively by direct manipulation of running instances with collaborative and integrative tools~\cite{amaraldemoaamas}. The IDE uses most of the \textit{endpoints} provided by \textit{jacamo-rest}.

\begin{figure}[ht]
\centering
\includegraphics[width=0.6\textwidth]{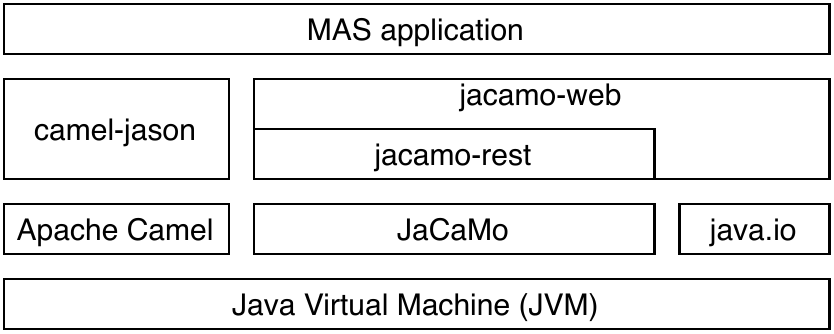}
\caption{Architecture of an example of application using \textit{jacamo-rest}}
\label{fig:rest-avaliacao}
\end{figure}

Figure~\ref{fig:rest-avaliacao} illustrates an integration architecture, which in this case extends the project \textit{jacamo-rest} to provide the web-based user interface \textit{jacamo-web}. Besides inheriting \textit{jacamo-rest} functionalities, the extension adds support for file access through the \textit{java.io} library. This is necessary because \textit{jacamo-web} provides static files, like HMTL pages and JavaScript files, to the client that then renders the interface. \textit{Jacamo-web} also accesses files of the JaCaMo platform, such as agent source code files, which can be manipulated through the \textit{jacamo-web} user interface. 

In addition, we illustrate in Figure~\ref{fig:rest-avaliacao} an implementation of an MAS application on the top of the architecture. In this example, the agents need to request data from external entities, as discussed in Section~\ref{integration}, something that \textit{jacamo-rest} does not provide native support for. To address this issue, we have used the approach proposed by~\cite{Amaral2019}, adding an Apache Camel component for agents that allows specifying communication routes using an extensive range of communication components for different protocols.

In this sense, this application is taking advantage of management integration for providing interfaces for changing agent, artefact and organisation implementations. Engineers that are using this IDE may collaborate when developing the MAS interactively, introducing pieces of code to the MAS through \textit{jacamo-rest} facilities.

\section{Discussion}\label{discussion}

The specific instantiation of the more generic resource-oriented management and interaction integration design, the implementation of the \textit{jacamo-rest} API, as well as the embedding of \textit{jacamo-rest} into \textit{jacamo-web}, raised some intriguing questions. 
An interesting aspect when providing a RESTful design interface to an MAS is the implementation of an interface that enables the navigation of historical states (\textit{time travel}\footnote{Time-travelling functionality is a feature that is, in particular, proposed in the context of mainstream programming languages like JavaScript~\cite{10.1145/2950290.2983933}.}).
Clients that request models from an MAS server need to be able to check whether a model resource they retrieve reflects the same model version of the previously retrieved resource.
This can be achieved by controlling resource revisions and providing them using specific endpoints such as \code{/agents/\{agentuid\}/revisions/\{revisionuid\}}.

A further question, as mentioned in Subsection~\ref{integration}, is how to expose the RPC-like JaCaMo interfaces to an agent's mailbox as well as to its command interface in a more RESTful manner.
This can potentially be achieved by treating messages and commands as resources, which are created by a \code{POST} request of the client, who then receives a UID of the corresponding message or command resource.
Retrieving the resource with a \code{GET} request can provide the status of the message or command, \textit{i.e.}, a description of how the agent has processed the message so far.
However, from a pragmatic perspective, one could alternatively conclude that REST is not a suitable protocol for the message and command interface. Instead, one could implement these parts of the API using a protocol that has real-time support as a first-class concern, for example, a publish/subscribe or socket-based interface.

We have to mention that our approach can be seen as limited since it has no native facility for the MAS working as a \textit{client} on the Web. One may suggest that besides a \code{/inbox} endpoint we can provide an \code{/outbox} facility. In this sense, an agent may add a message to a queue that should be managed by a process which later sends it to some external entity. However, it would also be necessary to provide proper representations for external entities which is out of the scope of this paper. In this sense, since the \textit{camel-jason} component~\cite{Amaral2019} provides such representations and since it is compatible with REST, we propose it for outbound communication needs.

\section{Future Work}\label{future-work}
This paper presents a snapshot of a work in progress. The following research directions can be considered as promising to further advance the line of work. \newline
\noindent \emph{Implement Revisioning and Time-Travel Functionality.} \\
In Section~\ref{discussion}, we have proposed integration of RESTful resource provision by the server and version control of specification object state. While the integration of a version control system and \textit{jacamo-rest} is a feature of the \textit{jacamo-rest}-based \textit{jacamo-web} system, the implementation is not yet adhering to the proposed approach.
Also, a more thorough evaluation of useful revisioning and versioning functionality will likely yield additional features, such as the explicit versioning of agents (\textit{minor, major, patch}) to enable version-based compatibility checking.

\noindent \emph{Generalise the \textit{jacamo-rest} interface for other MAOP systems.} \\
While the \textit{jacamo-rest} implementation is MAOP platform-specific, other AOP platforms can be exposed through a similar interface.
To enable true interoperability, framework-agnostic interfaces can be designed and implemented, building a heterogeneous ecosystem of agents and resources that are integrated at both run-time and design-time.

\noindent \emph{Implement outbound facilities, as well as additional inbound facilities.} \\
We have proposed an \code{/outbox} facility for agents that serve as clients. Besides this, environmental and organisational artefacts can also be shared from both a local system's perspective and an external system's perspective. For these functionalities, additional representations need to be modelled, and suitable endpoints need to be implemented.

\noindent \emph{Add better support of HATEOAS.} \\
Besides providing the \code{OPTIONS} method for each endpoint, it is recommended to provide links for each relation the retrieved resource has. For instance, when retrieving data about an agent, the agent resource should link to each artefact the agent is observing, which helps to find the right location of the resource.

\section{Conclusion}\label{conclusion}
In this paper, we have presented \textit{jacamo-rest}, a system that enables a resource-oriented approach to provide abstractions on multi-agent systems and their agents, artefacts, and organisations.
In contrast to related works, the focus of \textit{jacamo-rest} is on the provision of \textit{management} abstractions on MAS, and not primarily on enabling resource-oriented \textit{interactions} between agents and other Web resources.
Consequently, the presented work can be considered a stepping stone towards the integration of multi-agent oriented programming and modern software engineering approaches, in which the automated and (somewhat) autonomous management of software artefacts is a key concern.

\bibliographystyle{sbc}
\bibliography{bibliography}

\end{document}